\documentclass[10pt,twocolumn,letterpaper]{article}

\usepackage{times}
\usepackage{epsfig}
\usepackage{graphicx}
\usepackage{amsmath}
\usepackage{amssymb}
\usepackage{siunitx}

\usepackage{array}
\usepackage{textcomp}
\usepackage{authblk}
\usepackage{dcolumn} %align numbers by decimal point
\usepackage{todonotes}

\newcolumntype{d}[1]{D{.}{.}{#1}}

\newcolumntype{C}[1]{>{\centering\let\newline\\\arraybackslash\hspace{0pt}}m{#1}}

\usepackage[pagebackref=true,breaklinks=true,letterpaper=true,colorlinks,bookmarks=false]{hyperref}

\begin{document}
%%%%%%%%% TITLE

\title{Foreign Object Detection and Quantification of Fat Content Using A Novel Multiplexing Electric Field Sensor}

%\author{Anonymous Submission}

\author[1]{Anne E. Rittscher}
\author[2]{Aslam Sulaimalebbe\,$^*$}
\author[3]{Yves Capdeboscq \thanks{Corresponding Authors: aslam@zedsen.com, yves.capdeboscq@maths.ox.ac.uk, jens.rittscher@eng.ox.ac.uk}}
\author[4]{Jens Rittscher\,$^*$}
\affil[1]{Oxford University Innovation, Buxton Court, 3 West Way, Oxford, OX2 0JB, UK}
\affil[2]{Zedsen Ltd., Chiswick Park, 566 Chiswick High Road, London, W4 5YA}
\affil[3]{Mathematical Institute, Andrew Wiles Building, University of Oxford, Oxford OX2 6GG, UK}
\affil[4]{Department of Engineering Science, University of Oxford, Parks Road, Oxford, OX1 3PJ, UK}

\maketitle
\thispagestyle{empty}

%%%%%%%%% ABSTRACT
\begin{abstract}

There is an ever growing need to ensure the quality of food and assess specific quality parameters in all the links of the food chain, ranging from processing, distribution and retail to preparing food. Various imaging and sensing technologies, including X-ray imaging, ultrasound, and near infrared reflectance spectroscopy have been applied to the problem. Cost and other constraints restrict the application of some of these technologies. In this study we test a novel Multiplexing Electric Field Sensor (MEFS), an approach that allows for a completely non-invasive and non-destructive testing approach. Our experiments demonstrate the reliable detection of certain foreign objects and provide evidence that this sensor technology has the capability of measuring fat content in minced meat. Given the fact that this technology can already be deployed at very low cost, low maintenance and in various different form factors, we conclude that this type of MEFS is an extremely promising technology for addressing specific food quality issues. 

\end{abstract}

%%%%%%%%% Introduction & problem statement
\section{Introduction}

Sensing and imaging technologies play a vital role in ensuring the quality and safety of food in all links of the food chain, ranging from processing, distribution and retail to preparing food. In the same way any clinical decision depends on multiple diagnostic tests, different technologies are required to probe the biological, chemical and physical composition of food. Here, imaging technologies play a major role, as they allow to identify potential physical hazards caused by foreign objects and provide information about the composition of the biological specimen. Capacitive sensing is a technology that will enable the design and manufacture of very low-cost solutions while allowing a very flexible deployment. In the future, it will be possible to integrate such sensors into scales and other devices. Evaluating the applicability of capacitive sensing in this domain is the primary aim of this study. 

Minced meat has been purposely selected for our experiments as it is a representative meat product that allows to perform relevant tests under more controlled conditions. First, we consider the identification of foreign objects, which is highly relevant to the industry. Trafialek {\it et al.} \cite{trafialek2016risk} highlights that the presence of foreign bodies in food remains one of the main problems in food industry. In many countries the occurrence of foreign bodies is the most common cause of detected defects in foods. The range of contaminants is very diverse and depends on the specific product. The term “foreign body” refers to any unwanted objects in food, even if it comes from the same product (i.e. meat products containing bones, fruit products containing seeds) \cite{graves1998approaches}. Foreign materials in foods (glass, plastic, metal, etc.) are the biggest source of customer complaints received by many food manufacturers, retailers and enforcement authorities \cite{edwards2007observations}.

Thereafter, we designed a series of experiments that test the ability of measuring the fat content. Not only do consumers associate fat content in minced meat with quality, if qualified by words such as 'lean' or 'extra lean' the fat content needs to be within certain limits. Under annex VI, part B of Regulation (EU) 1169/2011 on the provision of food information to consumers minced meat may not be labelled with any of the descriptions in Table \ref{tab:minced-meat} unless it complies with the relevant criteria (being less than or equal to the number given, with the composition being 'checked on the basis of a daily average'). 

%\todo[inline]{Include table for fat content labels 
% https://www.businesscompanion.info/en/quick-guides/food-and-drink/composition-of-%meat-products#Mincedmeat
%}

\begin{table}[t]
\begin{center}
\small{
\begin{tabular}{ |l|c|c| } 
\hline 
 & F & C  \\ \hline
 lean minced meat & $\leq$ 7 \%  & $\leq$ 12 \% \\ 
 minced pure beef & $\leq$ 20 \% & $\leq$ 15 \%\\ 
 minced meat containing pigmeat & $\leq$ 30 \% & $\leq$ 18 \% \\
 minced meat of other species & $\leq$ 25 \% & $\leq$ 15 \% \\ 
 \hline 
\end{tabular} }
\end{center}
\caption{\textbf{Specific requirements concerning the designation of 
minced meat.} \it Table reproduced from regulation (EU) No 1169/2011 of the
European Parliament and of the Council (25 October 2011), Annex VI. 
Column F specifies fat content and column C specifies collagen/meat 
protein ratio.}
\label{tab:minced-meat}
\end{table}

Before we describe these experiments we provide an introduction to capacitive sensing (Section~\ref{s:capacitive-senors}) and compare it with related imaging technologies that are being used in the food industry in 
Section~\ref{s:related-technologies}. Materials and methods that are relevant to our experiments are given in 
Section~\ref{s:materials-and-methods}. The experimental results are presented in 
Section~\ref{s:experimental-results}. Finally, we summarise the contributions of this work and discuss conclusions in Section~\ref{s:summary}.

\section{Related Technologies}
\label{s:related-technologies}

A broad range of non-invasive techniques can be applied to monitor food quality \cite{butz2005recent}. While magnetic resonance imaging has been applied in the food industry \cite{mccarthy2012magnetic} it is a rather expensive technique, which is why we omit it from a more detailed review. X-ray and ultrasound imaging are already routinely used to detect foreign objects or contaminants. While less established, terahertz imaging provides some interesting opportunities. Near Infrared Reflectance Spectroscopy is a non-destructive methods that allows to assess the composition of various food products. Ideally, we are looking for a measurement technology that is non-destructive, non-invasive, repeatable, low cost, low maintenance, and easy to use.

\subsection{X-ray Imaging}

Haff and Toyofuku \cite{haff2008x} present a comprehensive review of applying X-ray inspection technology for the detection of defects and contaminants in the food industry. It is a well established technology  that allows
non-destructive imaging of interior features of a sample to detect hidden
defects or contaminants. The dangers of using ionising radiation, the necessary safety precautions, and relatively high costs are clear disadvantages. In addition to the potential harm caused to factory workers, the potential impact of ionising radiation with the tested specimens needs to be taken into account. 
X-ray inspection has the potential of detecting other foreign non-metallic material such as bone, glass, wood, plastic, and rocks. Mery and collaborators \cite{mery2011automated} present a system consisting of a X-ray source and flat panel detector for detecting bones in fish fillets. Overall, X-ray imaging can be used for certain food processing tasks in particular those in which packaging in cans, bottles, or jars need to be taken into account. Overall, the technique is quite costly and requires a specific setup. The need for analysis of the images provides additional challenges.

\subsection{Near Infrared Reflectance Spectroscopy}

Near infrared reflectance spectroscopy (NIRS) utilises the spectral range from 780 to 2500 nm and can provide provide complex structural information related to the vibration behaviour of combinations of molecular bonds. The energy absorption of organic molecules in NIRS region occurs when molecules vibrate or is translated into an absorption spectrum within the NIRS spectrometer. NIRS allows to generate a characteristic spectrum that behaves like a fingerprint for a given sample. NIRS is non-destructive and chemical-free method which can be easily used in continuous food quality evaluation \cite{williams1987near,cen2007theory}. 

Examples of applying NIRS techniques to specific applications in the food industry include the discrimination of edible oils and fats \cite{yang2005discriminant,hourant2000oil}, the detection of moisture in food \cite{wahlby2001nir}, and the evaluation of milk and dairy products \cite{cen2007theory}. Prieto {\it et al.} \cite{prieto2009application} present a review of applying NIRS techniques for predicting meat and mead product quality. The majority of the reviewed studies focused on predicting the major chemical components, some attempted to estimate the gross energy, myoglobin and collagen content in meat. NIRS has been applied to measure the fat, water and protein content of ground meat in a manufacturing setting \cite{togersen1999line}. 
 
Any quantitative analysis based on NIRS requires reliable calibration. Careful pre-processing of the acquired spectral data is necessary to account for background information and noise. In many practical settings robustness of calibration methods remain a concern and has to be applied regularly. While the cost of individual tests are low, the cost for instrumentation can be high.

\subsection{Ultrasound} 

Recent reviews \cite{knorr2004applications,khairi2015contact} provide a very comprehensive summary of applying ultrasonics in food processing. Potential applications include the inactivation of micro-organisms, ultrasonication of fruit juices, and the detection and identification of foreign bodies in packaged foods. In the meat sector, ultrasonic contact measurements had been used to for the characterisation of macroscopic properties,
such as the composition of raw meat mixtures \cite{benedito2001composition} or the textural properties \cite{llull2002evaluation}, and composition of fermented meat products \cite{simal2003ultrasonic}.

In most of these applications piezoelectric transducers are in direct contact with the specimen. In certain cases, the use of a couplant makes testing difficult, especially in the instances where food contamination is to be avoided. To address these limitations various groups have developed air-coupled non-contact systems that use capacitive devices
which are able to provide sufficient bandwidth for many measurements, including the detection of foreign bodies in a variety of food products including cheese, poultry and dough \cite{pallav2009air,cho2003foreign,corona2013advances}. Because non-contact ultrasound uses air as its medium, the transmitted signal is dependent on the conditions of the air, such as temperature, humidity, and airflow. Overall, ultrasound achieves a larger penetration capacity than NIR technique and has a lower cost and easiness of on-line implementation than NMR and X-rays systems.

\subsection{Teraherz Imaging}

Terahertz (THz) radiation, in the frequency range of 100 GHz to 10 THz (3mm to 0.03 mm), can pass through various objects, such as paper, vinyl, plastics, textiles, ceramics, semiconductors, lipids, and powders. Compared to X-ray imaging THz imaging can detect low density material  that are otherwise invisible with X-ray imaging. The ability of THz waves to pass through many dielectric packaging materials could facilitate nondestructive and noninvasive industrial inspections. Lee {\it et al.} \cite{lee2012detection} present results on detecting foreign objects including, pieces of aluminium, granite, maggots, and crickets in powder made from instant noodles. Ok and colleagues \cite{ok2014high} present efforts towards developing a low-cost, compact, simple, and fast detection system that could be used for industrial food inspection systems. However, the use of THz imaging is very limited \cite{armstrong2012truth}. The terahertz signal will decrease in power to 0.0000002 percent of its original strength after travelling just 1 mm in saline solution. Hence it can only be used for for surface imaging of things like skin cancer and tooth decay and laboratory tests on thin tissue samples.

\begin{figure*}[t]
	\begin{center}
		\includegraphics[height=1.1in]{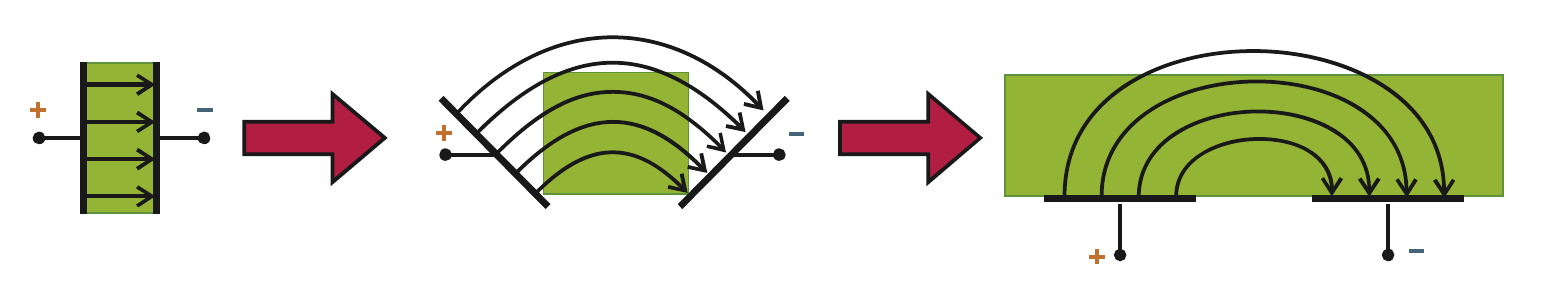}
	\end{center}
	\caption{\textbf{Parallel plate capacitor.} \it A parallel plate capacitor is 'folded up' to form a coplanar plate capacitor. The electric field lines now penetrate into the space above and the plates. The electric object is shown in green. The penetration depth is tuned by the voltage. }
	\label{fig:coplanar-capacitor}
	%\end{center}
\end{figure*}

\begin{figure}[t]
	\begin{center}
		\includegraphics[height=1.8in]{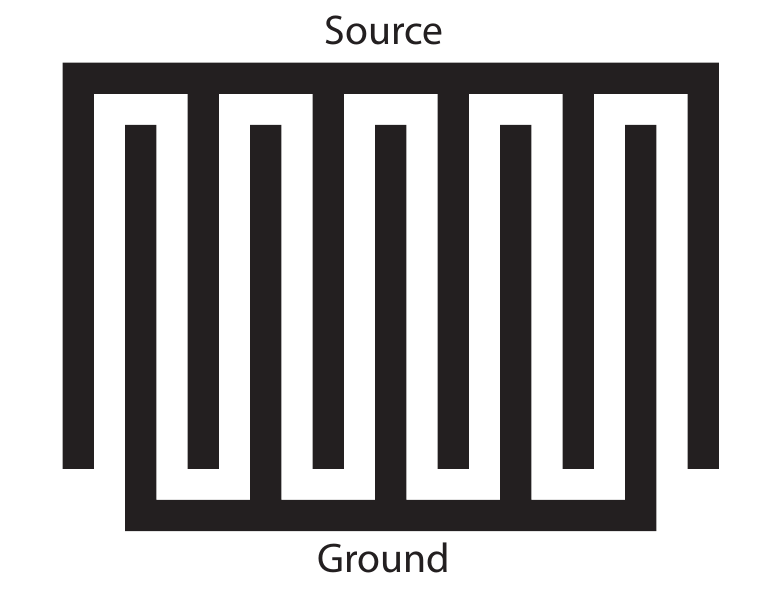}
	\end{center}
	\caption{\textbf{Interdigital sensor structure.} \it In most existing designs the electrodes follow a finger-like or digit-like pattern to implement a parallel plate capacitor.}
	\label{fig:interdigital-principle}
	%\end{center}
\end{figure}

\section{Capacitive Sensors}
\label{s:capacitive-senors}

The dielectric properties of materials \cite{solymar2009electrical}, in particular biological materials \cite{pethig1985dielectric,pethig1987dielectric}, are well understood. The dielectric properties of a given material provides valuable information about the storage and dissipation of electric and magnetic fields in the material. The polarizability of the material is expressed by permittivity. The permittivity is a complex number and the real part is often called as dielectric constant. Permittivity is typically measured as a function of frequency and are called dielectric/ impedance spectroscopy. The permittivity values show the interaction of an external field with the electric dipole moment of the sample. 

Interdigital capacitors \cite{mamishev2004interdigital} (IDCs) are used for the evaluation of near-surface properties, such as conductivity, permeability, and permittivity of materials. The interdigital capacitor sensor as shown in 
Figure~\ref{fig:interdigital-principle}, has the same principle of operation as the parallel plate or coaxial cylinder permittivity sensor 
\cite{mamishev2004interdigital}. The voltage is applied to the interdigital capacitor electrodes, and the impedance across the capacitor electrodes is changed due to the frequency and capacitance variations. Importantly, the parallel plate capacitor sensor, as shown in Figure \ref{fig:coplanar-capacitor}, does not require two-sided access to the specimen. Originally, developed in 1970 \cite{alley1970interdigital}, the design and layout of these devices has evolved dramatically. Since then, various different mathematical approaches have been developed to model the interdigital electrodes capacitance for multi-layered structures

Today, IDCs are widely used in chemical and biological sensing. For example, previous studies have demonstrated the successful application of IDCs to measuring sugar content \cite{angkawisittpan2012determination} and the dielectric properties of different fruits  \cite{komarov2005permittivity}. In the medical setting Hardick {\it et al.} \cite{hardick-brest-cancer} applied the technology to characterising breast cancer tissue. Mukhopadhyay and Gooneratne \cite{mukhopadhyay2007novel} use of the planar interdigital sensor to measure the fat content of different pork samples. The impedance measurements were compared to results obtained by a chemical analysis. Unfortunately, their study only contains 13 samples, two of which are being used for calibration. Their results provide some indication that IDCs, respond well to different percentages of fat content in pork. It should be noted that the results depend on the ambient temperature. 

\begin{figure}[t]
	\begin{center}
		\includegraphics[height=1.8in]{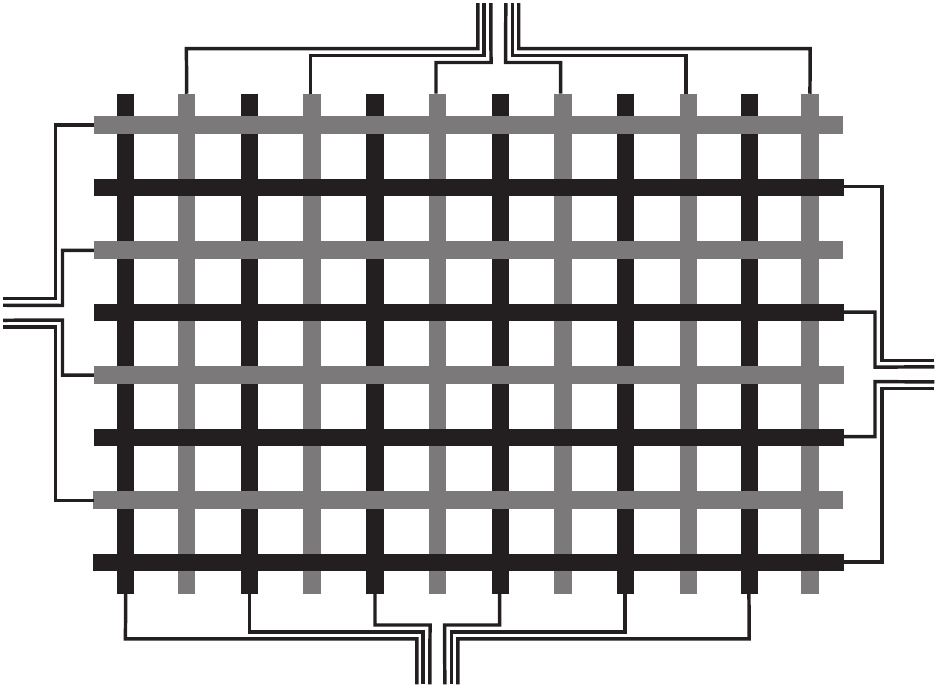}
	\end{center}
	\caption{\textbf{Zedsen Multiplexing Electric Field Sensor.} \it In contrast to traditional IDCs, Zedsen utilises a layout of horizontal and vertical electrode. In addition, to measuring traditional dielectric properties this sensor design allows to exploit complex permittivity in 3D space, i.e. on an $x$-$y$ grid and in the $z$-direction as penetration depth. }
	\label{fig:zedsen-sensor}
	%\end{center}
\end{figure}

\section{Multiplexing Electric Field Sensor (MEFS)}

The MEFS developed by Zedsen allows to exploit complex permittivity in 3D space, i.e. on a $x-y$ grid and in the $z$-direction as penetration depth as shown in Figure \ref{fig:zedsen-sensor}. Depending on the application, a penetration depth of several centimetres can be achieved by optimising the membrane layout, voltage, and operating frequency. By using a tomographic algorithm, the analysed object can be 'sliced' into layers, allowing to look non-invasively and non-distructively into the object. All required components are simple, cheap, versatile, and robust. 

%\begin{figure}[t]
%	\begin{center}
%		\includegraphics[width=0.4\columwnwidth]{figures/coplanar.eps}
%	\end{center}
%	\caption{A co-planar capacitive sensor is an unfolded parallel plane capacitor. The %dielectric is shown in green.}
%	\label{fig:coplanar-capacitor}
%	%\end{center}
%\end{figure}

\begin{figure}[t]
	\begin{center}
	    \begin{tabular}{cc}
		\includegraphics[height=1.2in]{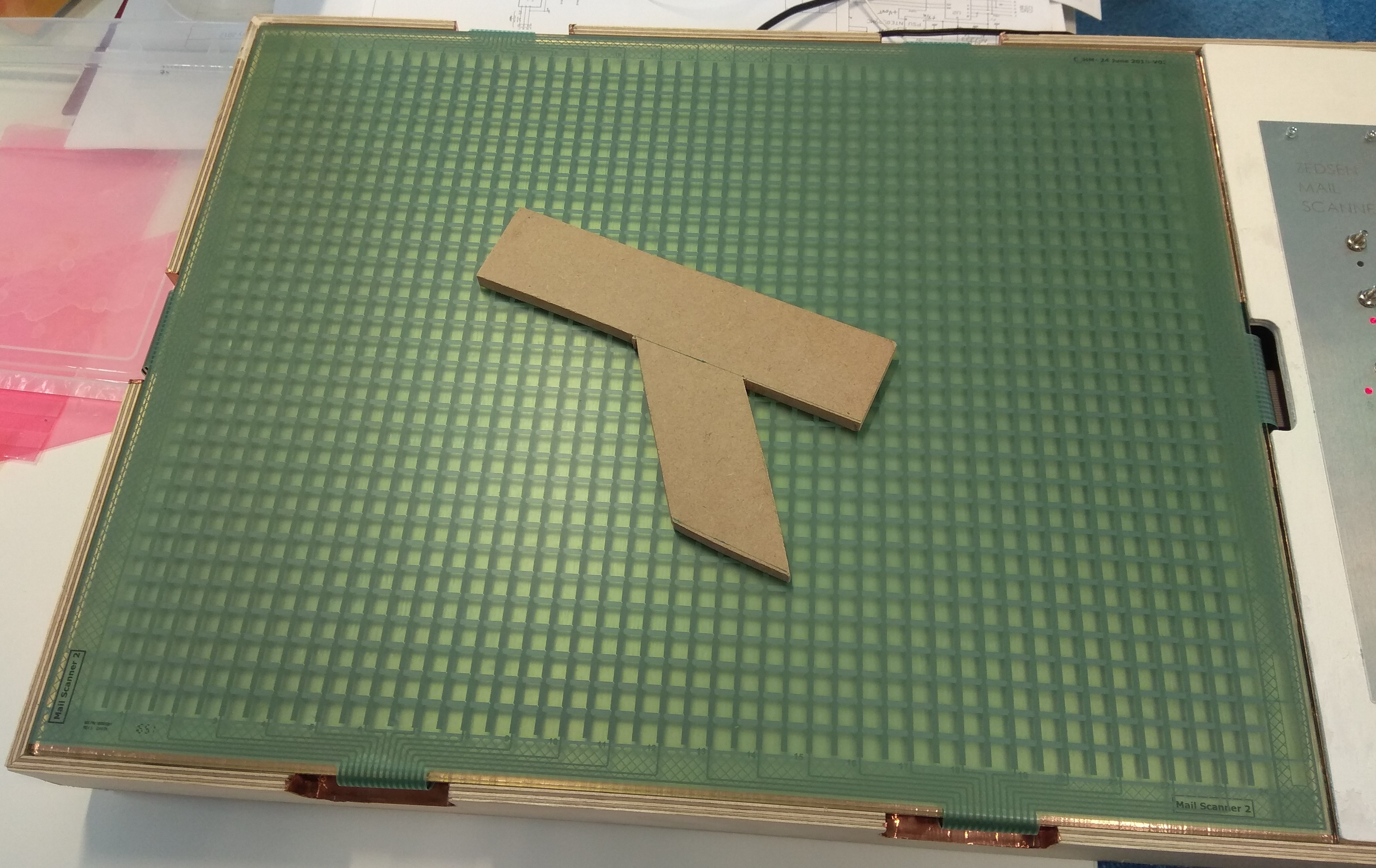} &
		\includegraphics[height=1.2in]{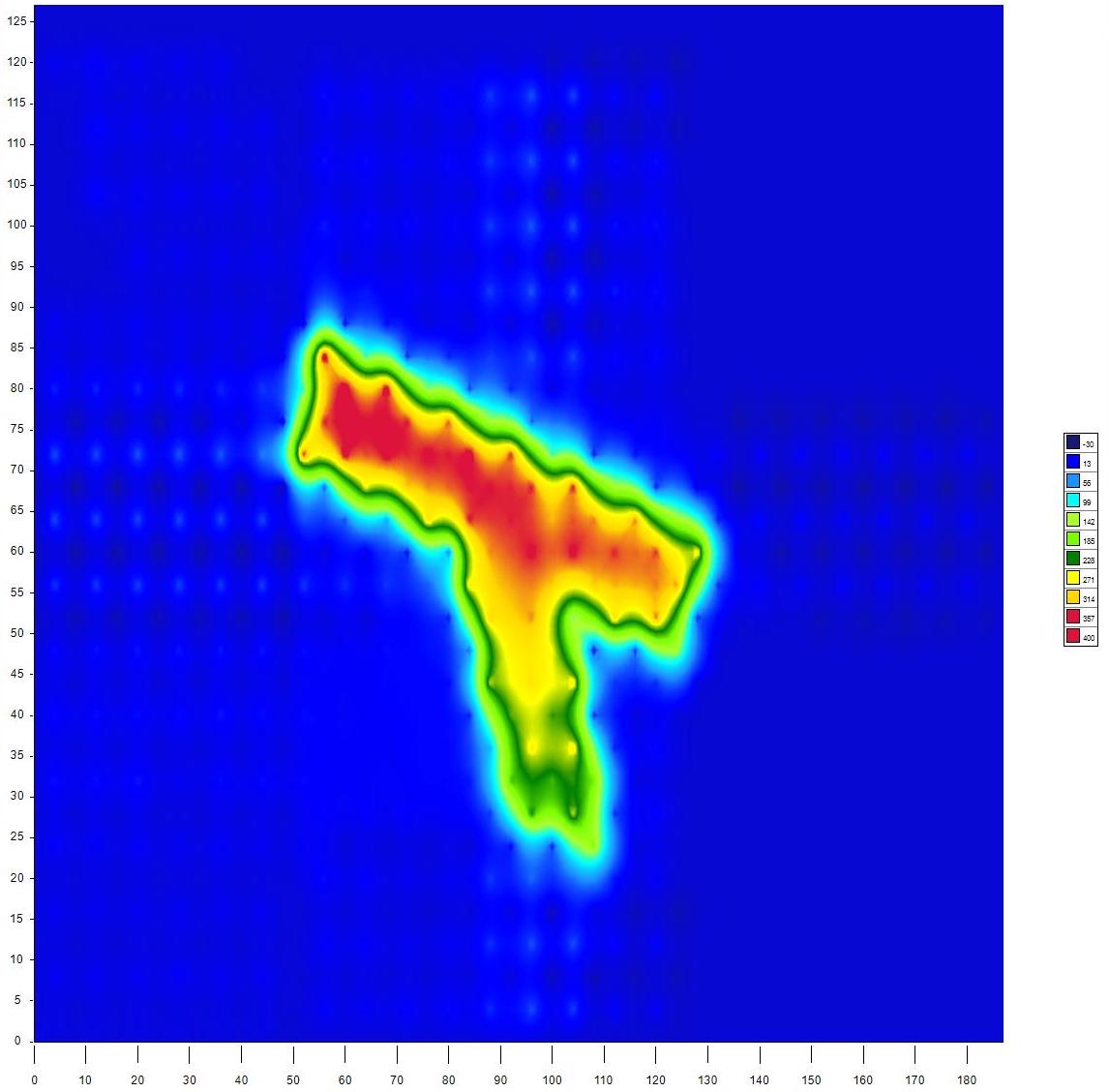} \\
		\end{tabular}
	\end{center}
	\caption{\textbf{Sample image.} \it This image of a wood cut provides an illustration of visualising dielectric properties in the spatial context of the object.}
	\label{fig:zedsen-image}
	%\end{center}
\end{figure}

The design builds on the principle of co-planar plate capacitor illustrated in Figure~\ref{fig:coplanar-capacitor}. In a parallel plate capacitor the electric field emerges perpendicular from the plate's surface. The penetration depth can be tuned by the voltage. Zedsen uses voltage pulses applied to the electrodes to generate an alternating electric field in the material to be probed.  Its electric permittivity $\epsilon$ then becomes a function of the oscillation frequency $\omega$. The basic structure of Zedsen MEFS is similar to the coplanar plate capacitor as illustrated in Figure  \ref{fig:coplanar-capacitor}. Using a cut out made from wood, Figure~\ref{fig:zedsen-image} illustrates the spatial localisation of dielectric measurements.

Crucially, several coplanar electrodes are used in order to measure an object with high resolution (see Figure~\ref{fig:zedsen-image}). The electrodes can be arranged in a parallel or grid fashion depending on the requirements of the application. The electrodes are printed on two sides of a transparent membrane using conductive ink containing, e.g., indium-tin or graphene. A dielectric layer (glass, polyamide, or polyester) covers the membrane on the top and an electrically grounded bottom layer is used to shield the device from the influence of stray capacities.

The permittivity for air is $\approx 1$ at \SI{0}{\celsius}, and water and porcelain have values of $\approx 80$ and $\approx 4$, respectively (at \SI{20}{\celsius}) \cite{levitskaya2000laboratory}. In other words, the permittivity varies greatly across the materials spectrum and can therefore provide a fingerprint for the various materials and materials combinations. Importantly, these known reference values provide a way to calibrate the sensor automatically. A prototype of a MEFS as shown in Figure \ref{fig:zedsen-image} is being tested in this study. 

\begin{figure}[t]
	\begin{center}
		\includegraphics[height=2.4in]{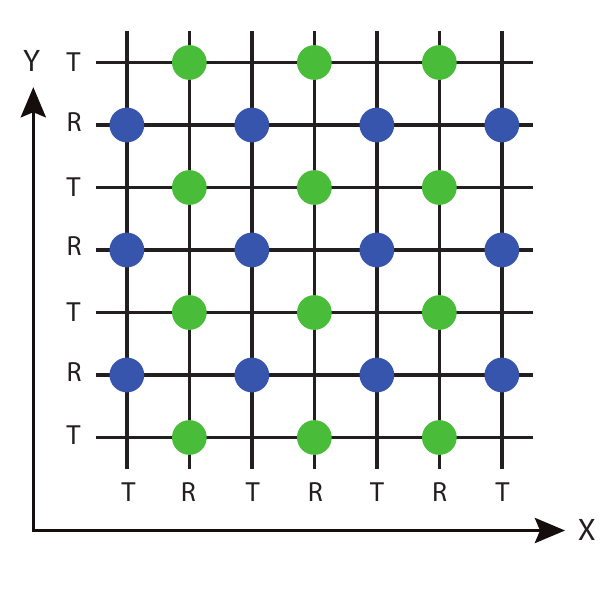}
	\end{center}
	\caption{\textbf{Cross-sectional measurements.} \it CSV (green) corresponds to $x$-transmitters
	and $y$-receivers whereas CSH (blue) corresponds to $y$-transmitters and $x$-receivers. CSV and CSH are combined to form a $xy$ matrix with the missing values linearly interpolated. }
	\label{fig:cross-sectional-measurements}
	%\end{center}
\end{figure}

%%%%%%%%% Description of the actual study 

\section{Materials and Methods}
\label{s:materials-and-methods}

Experiments were designed to test the capabilities of the specific MEFS sensor in a broad range of real word conditions. Rather than a well controlled laboratory environment, a domestic kitchen has been selected to conduct the described tests. For all experiments minced meat was wrapped in plastic film and was shaped in a block with the following dimensions: length: 10cm, width: 5cm, height: 2cm. 

To the test the presence of foreign objects, the object was placed on top of the block of minced meat. While one might argue that the object should be mixed into the minced meat, this experimental set up does in fact make the detection more difficult. For the purpose of measuring the fat content, a controlled mixture of lean beef and pure beef fat was used. Both were sourced from a local butcher. Before mincing the lean beef, all fat was trimmed. The mixed samples were measured at room temperature as well as cooled. A standard fridge/freezer was used to cool the mince between $3\,^{\circ}{\rm C}$ and $6\,^{\circ}{\rm C}$ degrees

\begin{figure}[t]
	\begin{center}
		\includegraphics[height=1.2in]{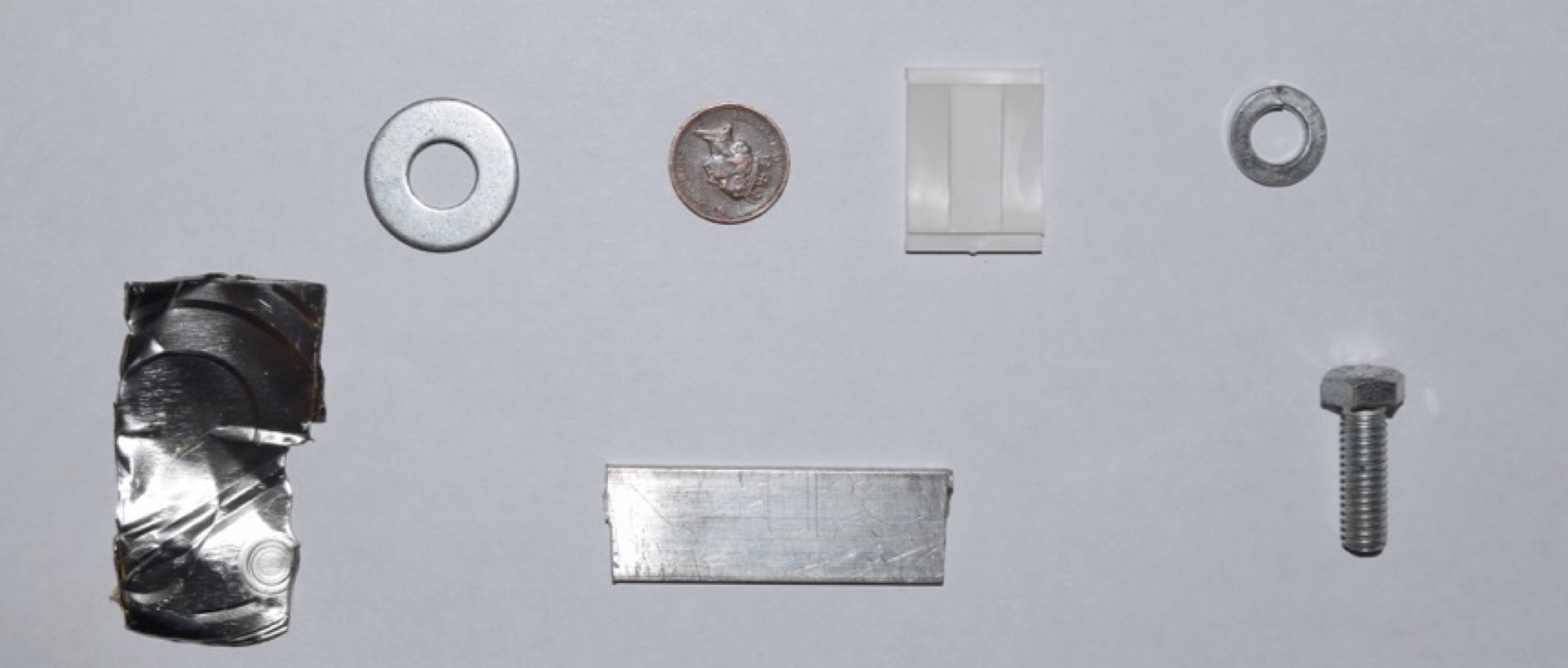}
	\end{center}
	\caption{\textbf{Collection of foreign objects.} \it Objects with a diameter of greater than 2 cm were selected for this study to account for the specific layout of the electrodes in the prototype sensor. See text for details.}
	\label{fig:object-collection}
	%\end{center}
\end{figure}

Here, we use the total output, referred to as $\gamma$ from the sensor. It corresponds to the voltage measured by cross directional measurements as illustrated in Figure \ref{fig:cross-sectional-measurements}: the emitting electrodee being orthogonal to the receiving one. The sums the output measured for all such cases  $2 \times n_v \times n_h$, where $n_v$ (resp. $n_h$) is the number of vertical (resp. horizontal) electrodee. In this prototype sensor the spacing between the vertical and horizontal electrodes is 1 cm. Shadowing and gain unbalance effects are removed by scaling the data by the "far-field", that is, the output measured away from the location of the sample. Although a more localised analysis would be possible, all measurements are based on the total output to characterise the measuring process and to evaluate the principal capabilities of this technology.

\section{Experimental Results}
\label{s:experimental-results}

As described in Section~\ref{s:materials-and-methods}, the volume and the geometry of the samples tested was controlled. 
Initial testing of the experimental setup demonstrated that the voltage output measured was strongly dependent on the temperature of the room, the temperature of the sample, the moisture buildup appearing between the sample and the sensor (for cooled samples). These observations correlate with the experiments reported by Mukhopadhyay and Gooneratne \cite{mukhopadhyay2007novel} discussed in Section \ref{s:capacitive-senors}. In addition, we also noticed an output buildup overtime, possibly due to moisture effects.

\subsection{Detection of Foreign Objects}
\label{s:experiments-foreign-objects}

\begin{figure*}[t]
	\begin{tabular}{cc}
		\includegraphics[height=1.6in]{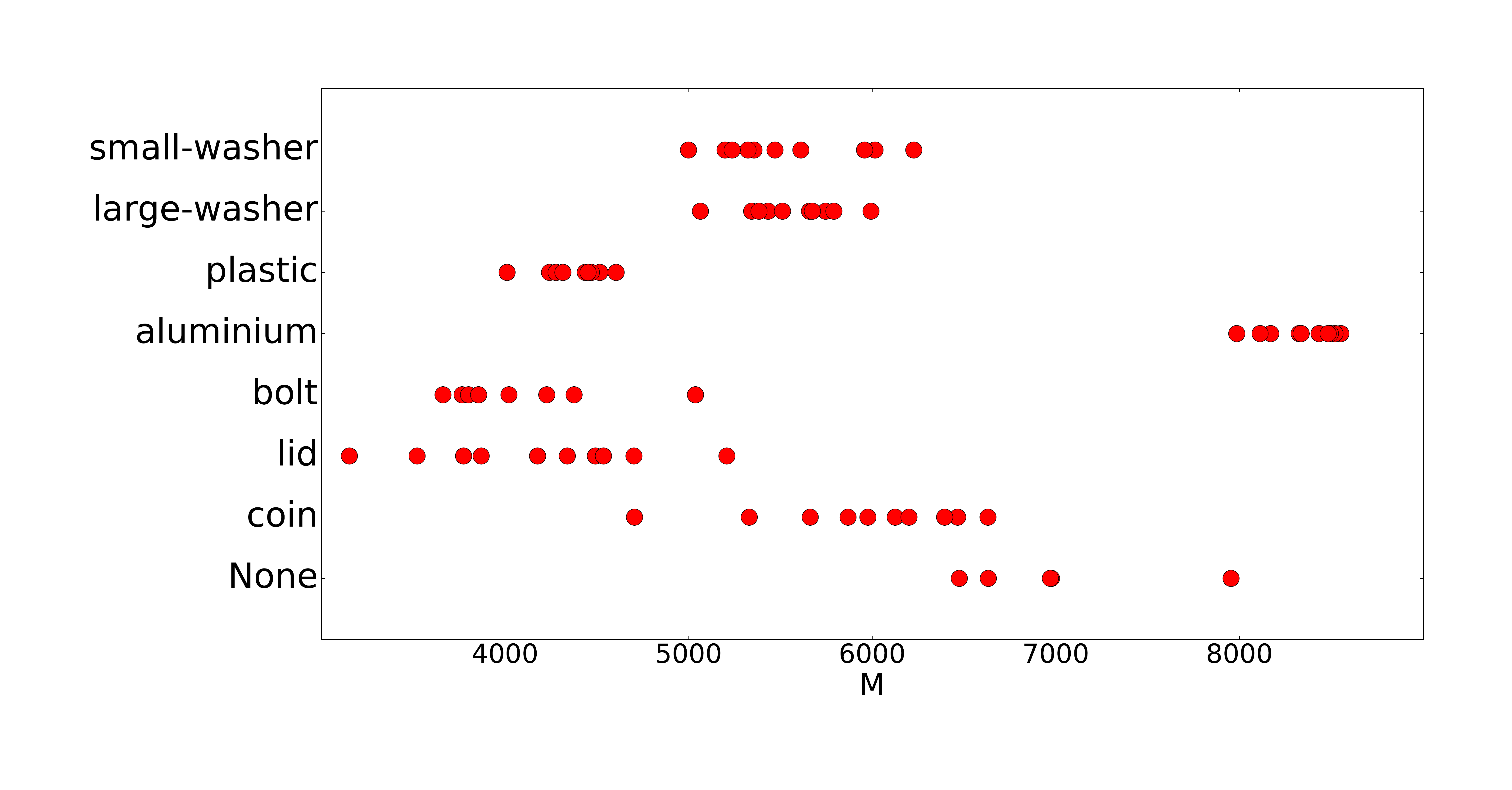} &
		\includegraphics[height=1.6in]{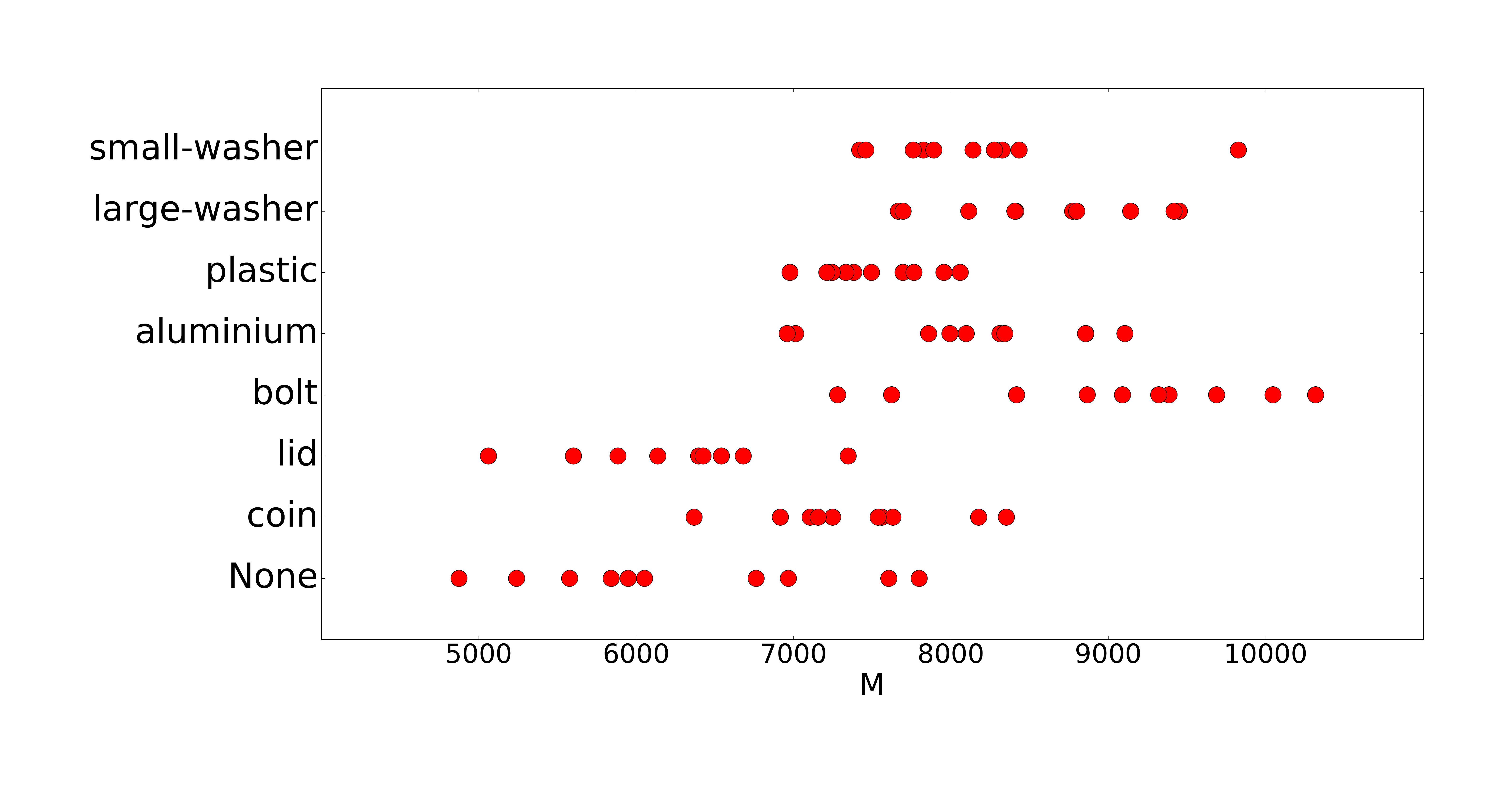} \\
		\textbf{\small 60 Volts} & \textbf{\small 100 Volts}\\
	\end{tabular}
	\caption{\textbf{Detection of foreign objects.} \it The graph summarises measurements of the total output when foreign objects were placed on top of the block of mined meat. In all these experiments the layer of mined meat was 2 cm thick.}
	\label{fig:measurement-foreign-objects}
	%\end{center}
\end{figure*}

A sample containing 50\% pure beef and 50\% beef fat was used of these experiments. The measurements, which were taken at room temperature are summarised in Figure~\ref{fig:measurement-foreign-objects}. Here, we selected objects that are about 2 cm in diameter to account for the 1 cm spacing of the electrodes. No additional processing of the total output was necessary. 

Our results indicate that changes in temperature, moisture and humidity do not affect the ability of detecting these foreign objects. While the did test working with frozen samples, we could not overcome the problem of moisture build up on the outside of the foil wrapping. It should be noted that if the objects would be mixed into the minced meat they would be easier to detect as they would be closer to the sensor. 

\subsection{Quantification of Fat Content}
\label{s:experiments-fat-content}

To factor out the dependence on temperature and moisture content, we present below the results obtained for 9 experiments scaled in terms of the estimated fat content. As motivated before we utilise the total output, $\omega$ of the sensor to test whether the sensor detects increases in fat content. For a given experiment, the sample containing pure beef is denoted by $S_0$, meaning that it contains $0\%$ fat. Similarly, we use the variable $S_{100}$ to refer to the sample which contains pure beef fat. To test the sensitivity of the sensor, we are preparing additional samples which contain $25\%$, $50\%$, and $75\%$ fat. The scaled output of the sensor is calculated as
\begin{equation}
    \Gamma(S_X) := \frac{\gamma(S_X) - \gamma(S_0)}{\gamma(S_{100}) - \gamma(S_0)} \,\,,
    \label{eq:scaled-output}
\end{equation}
with $X \in \{25, 50, 75\}$. Trials A,B,C and D where on a warm day at room temperature, with samples taken out of the fridge just before measurement, and the sensor left running in between. Trial E,F,G where made at a fixed room temperature of 20 degrees,and the samples were not cooled. Trial H,I,J where made at a fixed room temperature, with the samples kept between $3$ and $6$ degrees.

While these are rough measurements, they clearly document that the sensor detects increases in fat content.  Samples with larger fat content tended to warm up faster than samples with lower fat content. The table shows that the sensor orders correctly the samples in most cases. There is no linear relation between the output and the fat content, and nor should there be one. 

In total, 5 out of 50, or  10\% of the measurements are out of place. Namely these are: A: $\Gamma(S_{75})$, D: $\Gamma(S_{50})$, E: $\Gamma(S_{25})$, G: $\Gamma(S_{75})$ and  J: $\Gamma(S_{75})$.  In all experiments, no fat ($S_0$), full fat ($S_{100}$) and half fat ($S_{50}$) were correctly positioned.
The negative value E: $\Gamma(S_{25})$ is caused be the inhomogeneity of the sample and it is not a measurement error. It is likely that the fat and meat did not warm up to room temperature at the same time after cooling. 

%in terms of the estimated fat content. $0\%$ means minced beef (from the local butcher) where as $100\%$ beef fat (also from the butcher), whereas $X\%$ means the sample contains a proportion $1-X$ of minced beef, and a proportion $X$ of fat. These are rough measurements : the purpose is merely to document whether the sensor detects increases in fat content.  Trials A,B,C and D where on a warm day at room temperature, with samples taken out of the fridge just before measurement, and the sensor left running in between. Trial E,F,G where made at a fixed room temperature of 20 degrees,and the samples were not cooled. Trial H,I,J where made at a fixed room temperature, with the samples kept between $3$ and $6$ degrees. Samples with larger fat content tended to warm up faster than samples with lower fat content. The numbers given are, in each column) 
%\begin{center}
%    (total output(fat=$X\%$) - total output(fat=$0\%$))/((total %output(fat=$100\%$) - total output(fat=0\%)))
%\end{center}
%The total output corresponds to the voltage measured by cross directional measurements: the emitting antennae being 
%orthogonal to the receiving one. The output sums the output measured for all such cases  $2 \times n_v \times n_h$
%where $n_v$ (resp. $n_h$) is the number of vertical (resp. horizontal) antennae. %Shaddowing and gain unbalance effects are removed by scaling the data by the %"far-field", that is, the output measured away from the location of the sample. 

\begin{table}[t]
\begin{center}
\begin{tabular}{|c|c|c|c|c|c|}
\hline 
 & $\Gamma(S_{0})$ & $\Gamma(S_{25})$ & $\Gamma(S_{50})$ & $\Gamma(S_{75})$ & $\Gamma(S_{100})$\tabularnewline
\hline 
\hline 
A & 0 & 46 & 91 & 109 & 100\tabularnewline
\hline 
B & 0 & 36 & 49 & 75 & 100\tabularnewline
\hline 
C & 0 & 35 & 83 & 92 & 100\tabularnewline
\hline 
D & 0 & 4 & 62 & 57 & 100\tabularnewline
\hline 
E & 0 & -23 & 45 & 88 & 100\tabularnewline
\hline 
F & 0 & 24 & 75 & 78 & 100\tabularnewline
\hline 
G & 0 & 26 & 74 & 58 & 100\tabularnewline
\hline 
H & 0 & 44 & 59 & 77 & 100\tabularnewline
\hline 
I & 0 & 26 & 34 & 73 & 100\tabularnewline
\hline 
J & 0 & 69 & 85 & 83 & 100\tabularnewline
\hline 
\end{tabular}
\end{center}
\caption{\textbf{Quantification of fat content.} {\it The table shows the scaled total output measurement $\Omega$ as defined in Equation \ref{eq:scaled-output}. Notice, that in most cases the sensor has ordered the samples correctly with respect to fat content.}}
\label{t:fatcontent}
\end{table}

%The table shows that he sensor orders correctly the samples in most cases. There %is no linear relation between the output and the fat content, and nor should there %be one. 5/50 (10\%) of the measurements are out of place (A(75), D(50), E(25), %G(75) and  J(75)) but the bulk of the measurement reliably position the samples %correctly. In all experiments, no fat, full fat and half fat were correctly %positioned.

%%%%%%%%% Discussion and conclusion
\section{Summary \& Conclusion}
\label{s:summary}

These experimental results contribute demonstrate the promise of IDCs and MEFSs in various various applications in the food industry. Our measurements in Section~\ref{s:experiments-foreign-objects} demonstrate the reliable detection of certain foreign objects. Even taking environmental variables including humidity, moisture and changes of temperature into account, the sensor correctly positions the samples with respect to fat content. It should be noted that we only use the raw output of the sensor. No sophisticated calibration protocols are being applied to achieve these results. The fact that we achieve these results in a domestic kitchen, i.e. an uncontrolled environment, is an additional factor that should be considered. 

When compared to other related technologies reviewed in Section~\ref{s:related-technologies} several advantages need to be pointed out. When compared to X-ray imaging, IDCs do not produce any ionizing radiation, hence no specific shielding is required. For example, the sensor used in these experiments could be used on a counter in a retail environment or even at home. While near infrared reflectance spectroscopy does enable a more detailed biochemical characterisation of specimens it does require very careful calibration. Essentially, it is a laboratory technique that requires handling by a skilled operator. While ultrasound sensors and terahertz imaging are interesting in terms of technologies, they do require a very specific setup. This approach would certainly be justified, if food needs to be inspected as part of a fixed manufacturing process.  

Having demonstrated the principal applicability of the Zedsen sensor in terms of foreign object detection and the quantification of fat content, we would like to point out some of the specific advantages of this technology. The design of the sensor, in particular the number and spacing of the vertical and horizontal electrodee can easily be optimised to a specific application. Most importantly, this type of sensor can now be manufactured in various different forms, including thin flexible membranes. The technology itself is so versatile that it could be integrated into other devices, such as for example digital scales at very low cost. When compared to other measurement technologies it is truly passive and does not interfere with the tested specimens. Other sensors measuring the ambient temperature and humidity could easily be integrated to enable a more refined fully automated calibration process. In conclusion, this technology has the potential to be deployed in all the different links in the food chain, effectively addressing importance challenges of assessing food quality.

\section*{Acknowledgements} 

The sensor was provided by Zedsen Ltd. All measurements were generated by AER and JR outside the Zedsen laboratories. The study was sponsored by Zedsen. 

{\small
\bibliographystyle{apalike}
\bibliography{food-analysis-lit}

\begin{thebibliography}{}

\bibitem[Alley, 1970]{alley1970interdigital}
Alley, G.~D. (1970).
\newblock Interdigital capacitors and their application to lumped-element
  microwave integrated circuits.
\newblock {\em IEEE Transactions on Microwave Theory and Techniques},
  18(12):1028--1033.

\bibitem[Angkawisittpan and Manasri, 2012]{angkawisittpan2012determination}
Angkawisittpan, N. and Manasri, T. (2012).
\newblock Determination of sugar content in sugar solutions using interdigital
  capacitor sensor.
\newblock {\em Measurement Science Review}, 12(1):8--13.

\bibitem[Armstrong, 2012]{armstrong2012truth}
Armstrong, C.~M. (2012).
\newblock The truth about terahertz.
\newblock {\em IEEE Spectrum}, 49(9).

\bibitem[Benedito et~al., 2001]{benedito2001composition}
Benedito, J., Carcel, J., Rossello, C., and Mulet, A. (2001).
\newblock Composition assessment of raw meat mixtures using ultrasonics.
\newblock {\em Meat Science}, 57(4):365--370.

\bibitem[Butz et~al., 2005]{butz2005recent}
Butz, P., Hofmann, C., and Tauscher, B. (2005).
\newblock Recent developments in noninvasive techniques for fresh fruit and
  vegetable internal quality analysis.
\newblock {\em Journal of food science}, 70(9).

\bibitem[Cen and He, 2007]{cen2007theory}
Cen, H. and He, Y. (2007).
\newblock Theory and application of near infrared reflectance spectroscopy in
  determination of food quality.
\newblock {\em Trends in Food Science \& Technology}, 18(2):72--83.

\bibitem[Cho and Irudayaraj, 2003]{cho2003foreign}
Cho, B.-K. and Irudayaraj, J. (2003).
\newblock Foreign object and internal disorder detection in food materials
  using noncontact ultrasound imaging.
\newblock {\em Journal of Food Science}, 68(3):967--974.

\bibitem[Corona et~al., 2013]{corona2013advances}
Corona, E., Garcia-Perez, J.~V., Alvarez-Arenas, T. E.~G., Watson, N., Povey,
  M.~J., and Benedito, J. (2013).
\newblock Advances in the ultrasound characterization of dry-cured meat
  products.
\newblock {\em Journal of Food Engineering}, 119(3):464--470.

\bibitem[Edwards et~al., 2007]{edwards2007observations}
Edwards, M., Stringer, M., et~al. (2007).
\newblock Observations on patterns in foreign material investigations.
\newblock {\em Food control}, 18(7):773--782.

\bibitem[Graves et~al., 1998]{graves1998approaches}
Graves, M., Smith, A., and Batchelor, B. (1998).
\newblock Approaches to foreign body detection in foods.
\newblock {\em Trends in Food Science \& Technology}, 9(1):21--27.

\bibitem[Haff and Toyofuku, 2008]{haff2008x}
Haff, R.~P. and Toyofuku, N. (2008).
\newblock X-ray detection of defects and contaminants in the food industry.
\newblock {\em Sensing and Instrumentation for Food Quality and Safety},
  2(4):262--273.

\bibitem[Hourant et~al., 2000]{hourant2000oil}
Hourant, P., Baeten, V., Morales, M.~T., Meurens, M., and Aparicio, R. (2000).
\newblock Oil and fat classification by selected bands of near-infrared
  spectroscopy.
\newblock {\em Applied spectroscopy}, 54(8):1168--1174.

\bibitem[Khairi et~al., 2015]{khairi2015contact}
Khairi, M. T.~M., Ibrahim, S., Yunus, M. A.~M., and Faramarzi, M. (2015).
\newblock Contact and non-contact ultrasonic measurement in the food industry:
  a review.
\newblock {\em Measurement Science and Technology}, 27(1):012001.

\bibitem[Knorr et~al., 2004]{knorr2004applications}
Knorr, D., Zenker, M., Heinz, V., and Lee, D.-U. (2004).
\newblock Applications and potential of ultrasonics in food processing.
\newblock {\em Trends in Food Science \& Technology}, 15(5):261--266.

\bibitem[Komarov et~al., 2005]{komarov2005permittivity}
Komarov, V., Wang, S., and Tang, J. (2005).
\newblock Permittivity and measurements.
\newblock {\em Encyclopedia of RF and microwave engineering}.

\bibitem[Lee et~al., 2012]{lee2012detection}
Lee, Y.-K., Choi, S.-W., Han, S.-T., Woo, D.~H., and Chun, H.~S. (2012).
\newblock Detection of foreign bodies in foods using continuous wave terahertz
  imaging.
\newblock {\em Journal of food protection}, 75(1):179--183.

\bibitem[Levitskaya and Sternberg, 2000]{levitskaya2000laboratory}
Levitskaya, T.~M. and Sternberg, B.~K. (2000).
\newblock Laboratory measurement of material electrical properties: Extending
  the application of lumped-circuit equivalent models to 1 ghz.
\newblock {\em Radio Science}, 35(2):371--383.

\bibitem[Llull et~al., 2002]{llull2002evaluation}
Llull, P., Simal, S., Benedito, J., and Rossell{\'o}, C. (2002).
\newblock Evaluation of textural properties of a meat-based product
  (sobrassada) using ultrasonic techniques.
\newblock {\em Journal of Food Engineering}, 53(3):279--285.

\bibitem[Mamishev et~al., 2004]{mamishev2004interdigital}
Mamishev, A.~V., Sundara-Rajan, K., Yang, F., Du, Y., and Zahn, M. (2004).
\newblock Interdigital sensors and transducers.
\newblock {\em Proceedings of the IEEE}, 92(5):808--845.

\bibitem[McCarthy, 2012]{mccarthy2012magnetic}
McCarthy, M.~J. (2012).
\newblock {\em Magnetic resonance imaging in foods}.
\newblock Springer Science \& Business Media.

\bibitem[Mery et~al., 2011]{mery2011automated}
Mery, D., Lillo, I., Loebel, H., Riffo, V., Soto, A., Cipriano, A., and
  Aguilera, J.~M. (2011).
\newblock Automated fish bone detection using x-ray imaging.
\newblock {\em Journal of Food Engineering}, 105(3):485--492.

\bibitem[Mukhopadhyay and Gooneratne, 2007]{mukhopadhyay2007novel}
Mukhopadhyay, S.~C. and Gooneratne, C.~P. (2007).
\newblock A novel planar-type biosensor for noninvasive meat inspection.
\newblock {\em IEEE Sensors Journal}, 7(9):1340--1346.

\bibitem[Ok et~al., 2014]{ok2014high}
Ok, G., Park, K., Kim, H.~J., Chun, H.~S., and Choi, S.-W. (2014).
\newblock High-speed terahertz imaging toward food quality inspection.
\newblock {\em Applied optics}, 53(7):1406--1412.

\bibitem[Pallav et~al., 2009]{pallav2009air}
Pallav, P., Hutchins, D.~A., and Gan, T. (2009).
\newblock Air-coupled ultrasonic evaluation of food materials.
\newblock {\em Ultrasonics}, 49(2):244--253.

\bibitem[Pandya et~al., 2015]{hardick-brest-cancer}
Pandya, H.~J., Park, K., and Desai, J.~P. (2015).
\newblock Design and fabrication of a flexible mems-based electro-mechanical
  sensor array for breast cancer diagnosis.
\newblock {\em Journal of Micromechanics and Microengineering}, 25(7):075025.

\bibitem[Pethig, 1985]{pethig1985dielectric}
Pethig, R. (1985).
\newblock Dielectric and electrical properties of biological materials.
\newblock {\em Journal of Bioelectricity}, 4(2):vii--ix.

\bibitem[Pethig, 1987]{pethig1987dielectric}
Pethig, R. (1987).
\newblock Dielectric properties of body tissues.
\newblock {\em Clinical Physics and Physiological Measurement}, 8(4A):5.

\bibitem[Prieto et~al., 2009]{prieto2009application}
Prieto, N., Roehe, R., Lavin, P., Batten, G., and Andres, S. (2009).
\newblock Application of near infrared reflectance spectroscopy to predict meat
  and meat products quality: A review.
\newblock {\em Meat Science}, 83(2):175--186.

\bibitem[Simal et~al., 2003]{simal2003ultrasonic}
Simal, S., Benedito, J., Clemente, G., Femenia, A., and Rossell{\'o}, C.
  (2003).
\newblock Ultrasonic determination of the composition of a meat-based product.
\newblock {\em Journal of Food Engineering}, 58(3):253--257.

\bibitem[Solymar and Walsh, 2009]{solymar2009electrical}
Solymar, L. and Walsh, D. (2009).
\newblock {\em Electrical properties of materials}.
\newblock OUP Oxford.

\bibitem[T{\o}gersen et~al., 1999]{togersen1999line}
T{\o}gersen, G., Isaksson, T., Nilsen, B., Bakker, E., and Hildrum, K. (1999).
\newblock On-line nir analysis of fat, water and protein in industrial scale
  ground meat batches.
\newblock {\em Meat science}, 51(1):97--102.

\bibitem[Trafialek et~al., 2016]{trafialek2016risk}
Trafialek, J., Kaczmarek, S., and Kolanowski, W. (2016).
\newblock The risk analysis of metallic foreign bodies in food products.
\newblock {\em Journal of Food Quality}, 39(4):398--407.

\bibitem[W{\"a}hlby and Skj{\"o}ldebrand, 2001]{wahlby2001nir}
W{\"a}hlby, U. and Skj{\"o}ldebrand, C. (2001).
\newblock Nir-measurements of moisture changes in foods.
\newblock {\em Journal of Food Engineering}, 47(4):303--312.

\bibitem[Williams et~al., 1987]{williams1987near}
Williams, P., Norris, K., et~al. (1987).
\newblock {\em Near-infrared technology in the agricultural and food
  industries.}
\newblock American Association of Cereal Chemists, Inc.

\bibitem[Yang et~al., 2005]{yang2005discriminant}
Yang, H., Irudayaraj, J., and Paradkar, M.~M. (2005).
\newblock Discriminant analysis of edible oils and fats by ftir, ft-nir and
  ft-raman spectroscopy.
\newblock {\em Food Chemistry}, 93(1):25--32.

\end{thebibliography}
}

\end{document}